# Spectral Evidence for Amorphous Silicates in Least-processed CO Meteorites and Their Parent Bodies

Authors: Margaret M. McAdam[1, 2], Jessica M. Sunshine[1], Kieren T. Howard[3, 4], Conel M. O'D Alexander[5], Timothy J. McCoy[6], Schelte J. Bus[7]

[1] University of Maryland, Department of Astronomy, College Park, MD 20740. mmcadam@astro.umd.edu

[2] Northern Arizona University, Department of Physics and Astronomy, Flagstaff AZ 86011. maggie.mcadam@nau.edu

[3] American Museum of Natural History

[4] Kingsborough Community College

[5] Department of Terrestrial Magnetism, Carnegie Institution

[6] National Museum of Natural History, Smithsonian Institution.

[7] University of Hawaii, Institute for Astronomy

**Abstract**: Least-processed carbonaceous chondrites (carbonaceous chondrites that have experienced minimal aqueous alteration and thermal metamorphism) are characterized by their predominately amorphous iron-rich silicate interchondrule matrices and chondrule rims. This material is highly susceptible to destruction by the parent body processes of thermal metamorphism or aqueous alteration. The presence of abundant amorphous material in a meteorite indicates that the parent body, or at least a region of the parent body, experienced minimal processing since the time of accretion. The CO chemical group of carbonaceous chondrites has a significant number of these least-processed samples. We present visible/near-infrared and mid-infrared spectra of eight least-processed CO meteorites (petrologic type 3.0-3.1). In the visible/near-infrared, these COs are characterized by a broad weak feature that was first observed by Cloutis et al. (2012) to be at 1.3-µm and attributed to iron-rich amorphous silicate matrix materials. This feature is observed to be centered at 1.4-µm for terrestrially unweathered, least-processed CO meteorites. At mid-infrared wavelengths, a 21-µm feature, consistent with Si-O vibrations of amorphous materials and glasses, is also present. The spectral features of iron-rich amorphous silicate matrix are absent in both the near- and mid-infrared spectra of higher metamorphic grade COs because this material has recrystallized as crystalline olivine. Furthermore, spectra of least-processed primitive meteorites from other chemical groups (CRs, MET 00426 and QUE 99177, and C2-ungrouped Acfer 094), also exhibit a 21-μm feature. Thus, we conclude that the 1.4- and 21-μm features are characteristic of primitive least-processed meteorites from all chemical groups of carbonaceous chondrites. Finally, we present an IRTF+SPeX observation of asteroid (93) Minerva that has spectral similarities in the visible/near-infrared to the least-processed CO carbonaceous chondrites. While Minerva is not the only CO-like asteroid (e.g., Burbine et al., 2001), Minerva is likely the least-processed CO-like asteroid observed to date.

# 1 Introduction

The CO meteorites, named after their type member Ornans, are a group of carbonaceous chondrites that are characterized by small chondrules that are set in a relatively FeO-rich fine-grained matrix (Rubin et al., 2010; McSween, 1977). Here, matrix is used to include both interchondrule matrix and fine-grained rims on chondrules. All CO chondrites contain abundant (a few percent) metal that is indicative of their formation under relatively reducing conditions. CO chondrites experienced a range of thermal metamorphism in their parent bodies due to internal heating driven by the decay of short-lived $^{26}$Al. This variation in metamorphism is reflected in their petrologic types that range from 3.0 to 3.8 (Scott and Jones, 1990). Least-processed CO meteorites (petrologic types 3.0-3.1) are meteorites in the CO chemical group that have experienced minimal parent body processing including thermal metamorphism (petrologic types 3.2-3.6) and aqueous alteration (petrologic types 1 and 2; Alexander et al., 2018; Rubin et al., 2007; Harju et al., 2014). These meteorites contain matrices with high abundances of amorphous material (e.g., Brearley, 1993), presolar grains (e.g., Nittler et al., 2013), and insoluble organic matter contents (Alexander et al., 2007, 2014). Until recently, only one CO 3.0, Allan Hills (ALH) 77307, had been identified (Brearley, 1993; Grossman and Brearley, 2005) and studied using infrared spectroscopy (Cloutis et al., 2012). As a result of annual searches conducted in Antarctica, many more least-processed COs have now been discovered, including at least one new CO 3.0, Dominion Range (DOM) 08006 (e.g., Alexander et al., 2014; Davidson et al., 2015). However, these new samples have not been extensively studied spectroscopically.

A characteristic trait of least-processed carbonaceous chondrites, including those in the CO chemical group, is the presence of up to 30 vol. % amorphous material (iron-bearing silicates that lack long-range order) in their matrices (e.g., Alexander et al., 2018). This amorphous material is thought to be the product of disequilibrium condensation from a gas (e.g., Brearley, 1993; Abreu and Brearley, 2010; Greshake, 1997; Leroux et al, 2015) and is very susceptible to modification by aqueous alteration and/or thermal metamorphism. The presence of this amorphous material in a meteorite therefore indicates that it came from an asteroid or a region of an asteroid that experienced minimal processing after accretion. These samples are thus likely to retain important chemical and geological information about events in the early Solar System.

Previous studies of ALH 77307 indicate that the amorphous component of its matrix is an iron-bearing silicate (Brearley, 1993; De Gregorio et al., 2016). Spectroscopically, the amorphous materials in least-processed CO matrices behave like other amorphous silicates, including impact and volcanic glasses, due to their structural similarity. They have been identified in the visible/near infrared by a broad band near 1.3-µm (Cloutis et al., 2012). Iron oxides (rust) can also be present in least-processed COs (e.g., Brearley and Jones, 1998; Grossman and Rubin, 2006), but generally their abundance is low (less than a few volume percent) and are likely created by terrestrial weathering. Iron hydroxides and gypsum can also be present, but they are almost certainly the products of terrestrial alteration (Velbel et al., 1990).

Here, we present visible/near-infrared and mid-infrared spectral data for eight least-processed CO meteorites (3.0-3.1) to determine if the near-infrared spectral features identified by Cloutis et al., (2012) are representative of CO meteorites with minimal terrestrial weathering and to identify any characteristic spectral features in the mid-infrared. To investigate the spectral contribution of the amorphous iron-bearing silicate matrix in the mid-infrared, a simple first-order spectral model is used to remove the crystalline components. A potentially related asteroid for the least-processed carbonaceous chondrites, (93) Minerva, is identified. Minerva's near-infrared spectral features indicate that it is may be related to least-processed carbonaceous chondrites. Minerva appears to be rich in amorphous material, indicating it may be one of the least-processed parent asteroids observed to date, while other asteroids related to COs (e.g., Burbine et al., 2001) are thermally evolved. We explore the observational requirements for follow-up studies in the mid-infrared, which would confirm this potential detection and provide additional constraints on the origin of the least-processed carbonaceous chondrites.

## 2. Background

### 2.1 The least-altered CO meteorites

The least-processed CO meteorites are characterized by abundant chondrules (48 vol.% on average; McSween, 1977; Rubin, 2010; Ebel et al, 2016; Brearley, 2006), and fine-grained, largely amorphous (e.g., crystal structure exists on scales of 15 angstroms or less; Brearley, 1993) matrices. These meteorites have experienced minimal heating, which is indicated by several factors including the structural grade of the insoluble organic matter (Bonal, et al., 2007, 2016; Alexander, et al., 2014) and the relatively high and homogenous chromium content in olivines (Grossman and Brearley, 2005; Davidson, et al., 2014). These COs can also have high presolar circumstellar silicate grain abundances (e.g., Nittler et al., 2013), which is another indication of minimal parent body processing.

The amorphous-rich matrices of these least-processed COs have been extensively studied in certain CO samples (e.g., Brearley, 1993) and other low-metamorphic grade carbonaceous chondrites (e.g., Abreu and Brearley, 2010; Greshake, 1997; Le Guillou and Brearley, 2014; Abreu, 2016). The matrices of least-processed carbonaceous chondrites are dominated by truly amorphous material with little or no long range ordering, but do include some crystalline silicate grains (Brearley, 1993; De Gregorio et al., 2016). The matrix in ALH 77307 also contains highly localized 'proto-phyllosilicates' regions with some ordering, but these regions could be the result of terrestrial weathering (Brearley, 1993).

**2.2 Disequilibrium Condensation of Amorphous Silicates**

Amorphous silicates can form by several processes, including disequilibrium condensation, quenching of silicate melts, particle irradiation of crystalline materials, impact shock, and early stages of aqueous alteration (e.g., Alexander et al., 1989; Abreu and Brearley, 2010; Chizmadia and Nuth, 2006; Chizmadia, 2007; Nuth et al., 2005). However, studies of least-processed CO meteorites indicate that most of these processes can be ruled out as the formation mechanism for the amorphous materials (e.g., Brearley, 1993) and it is generally thought to be the product of disequilibrium condensation from rapidly cooled gas. Meteorites that preserve amorphous materials formed through disequilibrium condensation include least-processed carbonaceous chondrites in the CO, CR and ungrouped samples (e.g., Brearley, 1993; Abreu and Brearley, 2010; Greshake, 1997), CMs Paris and Yamato-791198 where these phases are preserved during heterogeneous aqueous alteration (Leroux et al, 2015; Chizmadia and Brearley, 2008) and ordinary chondrites (e.g., Alexander et al., 1989) . Glass with Embedded Metal and Sulfide (GEMS) grains found in interplanetary dust particles are also interpreted to form through disequilibrium condensation (Keller and Messenger, 2011; Messenger et al, 2015) and it has been proposed that they were the precursors for the amorphous materials found in chondrite matrices (e.g., Chizmadia and Brearley, 2008; Le Guillou and Brearley, 2014; Leroux et al., 2015; Keller and Messenger, 2012).

Most of the amorphous materials found in the matrices of least-processed carbonaceous chondrites likely formed in the solar nebula when temperatures in the disk were high and could vaporize dust (Wooden et al., 2007) or during transient heating events such as chondrule formation (e.g., Jones et al., 2000; Desch and Connolly, 2002; Connolly and Jones, 2017). These amorphous materials are highly susceptible to destruction through the parent body processes of thermal metamorphism and aqueous alteration. This paper confines itself to the study of least-processed carbonaceous chondrites that have not experienced extensive aqueous alteration. Some low-metamorphic grade COs, including DOM 08006, may contain up to ~0.5 wt.% water (Alexander et al., 2018), as do some CRs (Alexander et al., 2013). However, unlike meteorites that experienced significant interactions with water (e.g., Paris and Yamato-791198; Leroux et al, 2015; Chizmadia and Brearley, 2008), the least-processed COs have relatively high abundances of amorphous-iron bearing materials that have been quantified by powder X-ray diffraction and TEM (e.g., Howard et al., 2015; Alexander et al., 2018), high pre-solar grain abundances (Nittler et al., 2013) and are highly chemically unequilibrated (e.g., Grossman and Brearley, 2005; Davidson et al., 2014). The meteorites in this study can, therefore, be considered both primitive and least-processed.

**2.3 Spectroscopy of Amorphous Silicates and Glasses**

Glasses are known to form by rapid cooling of molten silicates in events such as impacts (e.g., Tompkins et al., 2010), volcanic eruptions (e.g., Crisp et al., 1990; Bell and Mao, 1976) and chondrule formation (e.g., Kurahashi et al., 2008). However, this is not the only way to make amorphous materials. Disequilibrium condensation and energetic particle irradiation can also produce amorphous materials, and these may be the mechanisms responsible for producing the amorphous materials that are associated with the 10- and 20-μm spectral features in the interstellar medium (e.g., Kemper et al, 2004, 2005).

The amorphous silicates found in the matrices of least-processed COs are similar to other silicates lacking long-range order, including glasses formed by impact and volcanism.

Structurally, glasses are comprised of units with short- to mid-range order (~15-20-Å; King et al., 2004). As discussed by King *et al.* (2004), when glass is formed, silicon and oxygen atoms form tetrahedra, locally. If there are other ions present, these enter the local structure as network formers or network modifiers. Network formers behave similarly to the silicon atoms, creating tetrahedra and increasing the ordering. These cations tend to be tri- or quadrivalent (*e.g.*, $Fe^{3+}$, $Ti^{4+}$, or $Al^{3+}$). Mono- or divalent ions (e.g., $Fe^{2+}$, $Mg^{2+}$, $Ca^{2+}$) modify the local ordering by forming bonds with oxygens, thus limiting structural ordering by preventing bond formation between tetrahedra.

In the mid-infrared, the cation-oxygen bonds produce vibrations between 8-12.5-μm, and 16-25-μm. The exact positions of these mid-infrared features depend on several factors including the degree of polymerization (or the number of cation-oxygen bonds), cation substitutions in the tetrahedra, the number of network modifiers present and, for iron-rich glasses, the redox conditions (Mysen, 1982; Mysen et al., 1982; McMillian, 1984a, b, Dorschner et al., 1995; Mysen and Richet, 2005). The 8-12.5-μm features are caused by asymmetric stretching vibrations of the O-Si-O/Si-O-Si bonds. Glasses can have complex features in this region because ordering only exists locally and different units can create multiple features (King et al., 2004; McMillian, 1984a, b). Additionally, there are two asymmetric stretching modes that also produce multiple features between 8-10-μm and ~12.5-μm. Furthermore, the substitution of cations can change these frequencies. Increasing polymerization of glasses, or the sharing of tetrahedrally coordinated oxygens, shifts spectral features to shorter wavelengths (from 9-μm to ~8-μm). Long-wavelength features (16-25-μm) are caused by the bending vibrations of the Si-O-Si bonds (King et al., 2004; Dorschner et al., 1995). Network formers other than silicon (e.g., aluminum) can change the frequencies of these vibrations (King et al., 2004; Dorschner et al., 1995).

In the visible/near-infrared, impact and volcanic glasses have well known crystal field features (e.g., Bell and Mao 1976; Moroz et al., 2009; Horgan et al., 2016; Morlok et al., 2016). Crystal field features are created through an electronic process where an ion with unfilled d-orbitals exists in a crystal structure. The bonding environment changes the energy levels of the d-orbitals, generally splitting them into two energy states (Burns, 1993). Depending on the bonding environment (octahedral vs. tetrahedral), these states can be excited to produce absorption features in the visible/near-infrared. While the effects of ferric/ferrous iron on the near-infrared and mid-infrared spectra of glasses has not been extensively examined (e.g., Horgan et al., 2016), it has been suggested that ferric iron causes a broad absorption over the visible/near-infrared wavelengths that causes lower albedos and a moderately red-slope (Moroz et al., 2009). If ferrous iron cations are in local octahedral coordination, they produce the crystal field transitions seen at ~1-µm in the spectra of natural and synthetic glasses (Moroz et al, 2009; Bell and Mao, 1976; Morris et al., 2000). Glasses with basaltic compositions have been extensively studied, but there is a paucity of studies that have investigated the simple iron/magnesium-silicon-oxygen system (e.g., King et al., 2004; Crisp et al., 1990).

**2.4 Previous Spectral Studies of CO meteorites**

Few studies (e.g., Cloutis et al., 2012) have documented the spectral characteristics of least-processed CO meteorites. This is largely due to the lack of least-processed COs in the meteorite collection. Previous spectral studies of COs mainly examined more metamorphosed samples such as Ornans (3.4), Warrenton (3.7), and Lancé (3.5) (e.g., Johnson and Fanale, 1973; Gaffey, 1976; Salisbury et al., 1975). These higher metamorphic grade COs are characterized in the near-infrared and mid-infrared by absorptions caused by the olivine and pyroxene present in their chondrules and matrix. The first CO3.0 to be identified, ALH 77307, is spectrally characterized in the near-infrared by a flat slope and a broad, weak ~1.3-μm feature that has been attributed to amorphous iron-bearing matrix silicates (Cloutis et al., 2012). In this wavelength region, ALH 77307 generally has a lower albedo than more metamorphosed COs (Cloutis et al., 2012).

Least-processed CO chondrites are mineralogically (e.g., Alexander et al., 2018) and, therefore, spectrally distinct from other groups of carbonaceous chondrites. The CM1/2, CI1, and CR1/2 carbonaceous chondrites all have abundant hydrated minerals (e.g., Howard et al., 2009, 2011, 2015; King et al., 2015) produced by aqueous alteration. Hydrated minerals dominate the spectra of these carbonaceous chondrites in both the visible/near-infrared and mid-infrared. The aqueously altered CMs and CIs can have a charge transfer band at 0.7-μm in the near-infrared, as well as strong features at 11.5-12.5-μm, 16-μm and 21-μm caused by the vibrational absorptions of the phyllosilicates. In the least altered CMs, which may contain ~25-30 vol.% olivine (Howard et al., 2009, 2011), a feature at 19.5-μm is also present in their spectra (Beck et al., 2014; McAdam et al., 2015a). Least-processed COs are also spectrally distinct from more thermally metamorphosed carbonaceous chondrites, including COs and CVs. During thermal metamorphism minerals in chondrules and matrix equilibrate, homogenizing the olivine composition and increasing the crystallinity of the silicates in the meteorite. The mid-infrared spectral features are therefore dominated by the abundant olivine (e.g., McAdam et al., 2015b). In the near-infrared, thermally metamorphosed meteorites exhibit olivine and pyroxene features at 1-μm and 2-μm. For meteorites with similar metamorphic grades as the COs studied here, there are both mineralogical and spectral similarities, which will be discussed further in *Section 4*.

## 3. Samples and Data Collection

### 3.1 Samples

The meteorites examined in this study are listed in **Table 1**, along with their metamorphic grades, weathering grades, and chemical group. The least-processed CO meteorites were selected since they appear to contain amorphous iron-bearing silicates in their matrices that comprise up to ~30 vol.% of the meteorite (e.g., Alexander et al., 2018). Approximately 100 mg of material is sub-sampled from larger chips of the meteorite. These 100 mg aliquots are first prepared for Position-Sensitive X-ray Diffraction analyses (PSD-XRD). PSD-XRD preparation involves grinding each aliquot to a fine powder (grain size of ≤35-μm, estimated with optical microscopy). Multiple aliquots from four meteorites were analyzed for comparison purposes. The samples, after grinding, are spectrally characterized.

All of the samples show at least some evidence of terrestrial weathering. On a scale from A, minimal weathering, to C, moderate weathering, most of the samples studied here have grades A, A/B or B, i.e., minor to moderate rust along fractures (including minor-moderate rust halos on metal grains). Included in our suite are two meteorites that have more terrestrial weathering (Miller Range (MIL) 090073 and 07687). These meteorites are reported to have moderate to severe rustiness, and in the case of MIL 07687 evaporite minerals were visible on its outer surface when it was found.

In previous analyses of these meteorites, organic solvents have been used successfully to remove terrestrial weathering for bulk isotope analyses (Greenwood et al., 2016). A similar protocol was not used here, since the organic solvents could add spurious spectral features to the

near- and mid-infrared spectral regions. Consequently, terrestrial weathering will affect the spectra presented here to some degree. These affects will be discussed in the following sections.

**Table 1: Meteorite Samples**

| Meteorite | # of aliquots | Chemical Group | Metamorphic Grade | Weathering Grade |
|---|---|---|---|---|
| ALH 77307 | 1 | CO | 3.0 | Ae[+] |
| DOM 08006 | 2 | CO | 3.0 | A/B |
| MIL 07687* | 1 | CO | 3.0 | Ce |
| DOM 08004 | 2 | CO | 3.1 | B |
| MIL 090010 | 1 | CO | 3.1 | A/B |
| MIL 07193 | 2 | CO | 3.1 | A |
| MIL 07709 | 2 | CO | 3.1 | A |
| MIL 090073 | 1 | CO | 3.1 | B/C |
| Felix | 1 | CO | 3.3/3.4 | - |
| ALH 85003 | 1 | CO | 3.5 | A/B |
| ALH 83108 | 1 | CO | 3.5 | A |
| MET 00426 | 1 | CR | 3 | B |
| QUE 99177 | 1 | CR | 3 | Be |
| Acfer 094 | 1 | Ungrouped | 3.0 | - |
| Allende | 1 | CV3 | 3.6 | - |
| Yamato-791198** | 1 | CM2 | - | - |

*This meteorite may not be a CO, but related to Acfer 094 (Brearley, 2012; Davidson et al., 2014).
[+]Here, 'e' indicates the visual presence of evaporites on the meteorite.
**Published in McAdam et al., 2015.

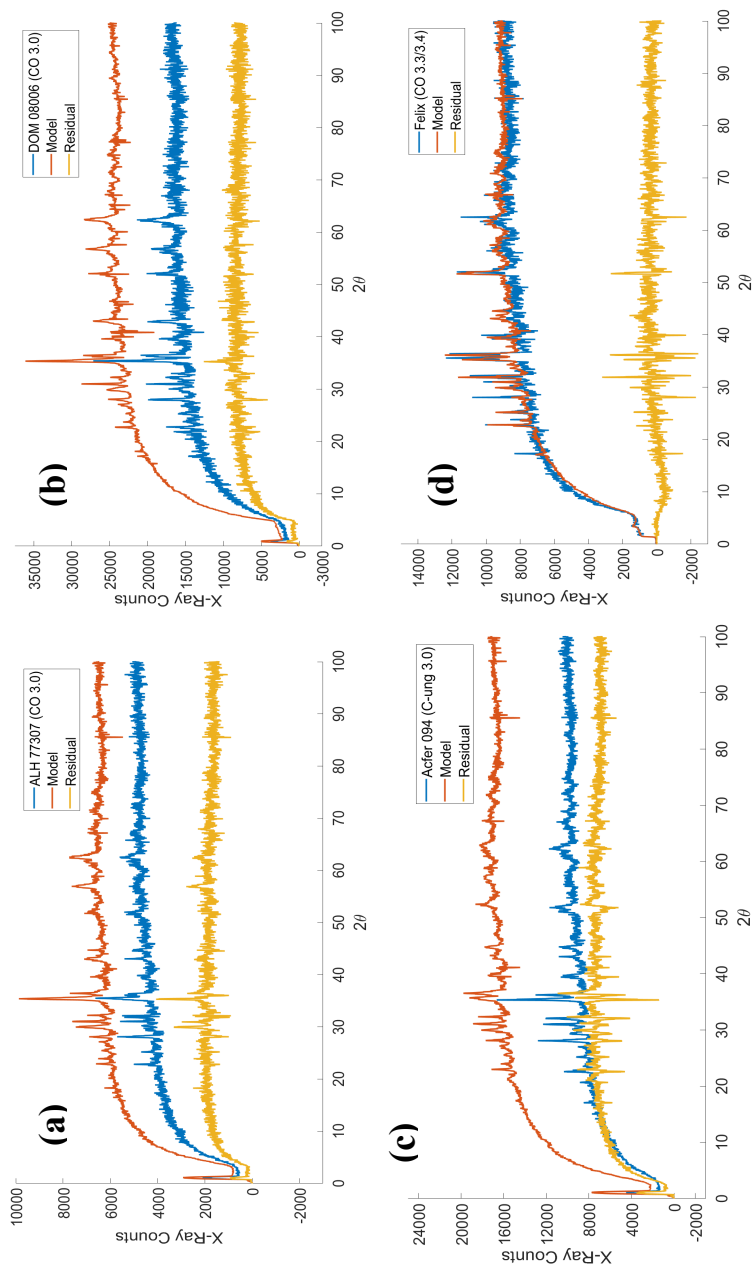

**Figure 1:** PSD-XRD Spectrum, model and residual for (a) ALH 77307 (CO 3.0), (b) DOM 08006 (CO 3.0), (c) Acfer 094 (C-ungrouped 3.0) and (d) Felix (CO 3.3/3.4). For ALH 77307, DOM 08006 and Acfer 094, crystallographic peaks in the X-ray patterns can be completely accounted for, however significant X-ray residuals remain. These X-ray counts are attributed to the poorly crystalline Fe-bearing matrix materials. Conversely, Felix has negligible amorphous materials; its XRD spectrum can be completely modeled with crystalline standards without significant residuals. Felix has been heated to a higher peak temperature than DOM 08006, and ALH 77307 so the amorphous matrix materials have been recrystallized.

## 3.2 Modal Mineralogy of Selected Samples

The powders used for the spectral analyses have previously been studied using position-sensitive X-ray diffraction to constrain modal mineral abundances following the approach described previously and summarized below (Howard et al. 2015; Alexander et al., 2018). Modal abundances for the suite of CO samples in this study have recently been reported in Alexander et al. (2018). Here we focus on the two lowest-metamorphic grade COs (ALH 77307, DOM 08006) and CO-related meteorite (Acfer 094), as well as one thermally processed CO meteorite, to illustrate the mineralogy of these samples and the changes associated with thermal metamorphism. The collected PSD-XRD patterns for these meteorites are shown in **Figure 1a-d**, along with phases identified using the International Centre for Diffraction Data (ICDD) database. To quantify the proportions of mineral phases in the meteorites, standards of pure minerals are measured under identical conditions. Then, peak intensities in the standard patterns are normalized for differences in collection times and scaled to match the meteorite patterns. Sequentially, standard patterns are subtracted from the meteorite pattern in identified proportions until a peak free residual is reached. Once the proportions of each standard are determined using the pattern fitting techniques, the pattern-proportions are converted into mineral proportions (wt. %) by correcting for relative difference in X-ray absorption using calculated mass absorption coefficients.

For ALH 77307, DOM 08006 and Acfer 094, the pattern fitting routines accurately account for all the crystalline peaks. However, significant X-ray residuals remain after subtraction of the crystalline features due to the fluorescence of Fe-bearing material without long-range crystallographic order. Both Fe-bearing silicates and Fe-oxides/hydroxides produced during terrestrial weathering are known to be X-ray amorphous in carbonaceous chondrites. The residual signature itself cannot be used to determine the identity of the contributing phases; only the total contribution of residual X-ray counts to the bulk diffraction pattern can be quantified. However, the samples used in the PSD-XRD and spectral analyses were carefully chosen from the interior regions of the meteorites to reduce terrestrial weathering. Furthermore, observations of these meteorites using optical microscopy and Transmission Electron Microscopy (TEM; e.g., Brearley, 1993) excludes significant contributions to the X-ray residuals from Fe-oxides/hydroxides since the abundances of these phases could not create the residuals observed. Amorphous silicates in chondrules (mesostases) tend to have low iron contents and will only contribute weakly to the X-ray residuals (Howard et al., 2015; Kurahashi et al, 2008). Sections of least-processed CO meteorites studied using TEM methods, including ALH 77307 and Acfer 094, confirm the presence of abundant and amorphous Fe-rich silicate in the matrices of these samples (Brearley, 1993, Greshake, 1997, De Gregorio et al., 2016). Previous studies have used bulk compositions and petrography to demonstrate that most amorphous material identified by PSD-XRD in primitive CR 3.0 samples is also Fe-rich silicate in matrix (e.g., Howard et al., 2015a).

    The resulting modal mineralogies are listed in **Table 2**. The CO meteorite Felix (**Figure 1d**) has experienced moderate metamorphism and is petrologically classified as 3.3/3.4. The petrologic grade of Felix is relatively low compared to other CO meteorites, such as ALH 85003 (CO 3.5), that experienced higher peak temperatures. PSD-XRD measurements indicate that unlike the least-processed COs, Felix's PSD-XRD spectrum can be completely modeled using crystalline components. There is negligible X-ray residual for this meteorite and, therefore, negligible amorphous iron-bearing matrix. The lack of amorphous components in the PSD-XRD measurements of Felix is consistent with petrographic studies of this meteorite that show that

even after mild to moderate metamorphism the amorphous silicate matrix materials have recrystallized to crystalline minerals.

**Table 2: Modal mineralogy for selected samples.**

| Meteorite | Type | Fo100 | Fo90 | Fo80 | Fo70 | Fo60 | Fo40 | Fo25 | Tot. Olv. | En. | ClinEn. | Tot. En. | Sulfide | Metal | Mag | Ph. | Am | Total |
|---|---|---|---|---|---|---|---|---|---|---|---|---|---|---|---|---|---|---|
| ALH 77307 | 3.00 | 0% | 21% | 0% | 0% | 0% | 14% | 0% | 35% | 27% | 0% | 27% | 3% | 2% | 8% | 4% | 20% | 100% |
| DOM 08006 | 3.00 | 42% | 0% | 0% | 0% | 10% | 0% | 0% | 42% | 22% | 0% | 22% | 3% | 1% | 6% | 2% | 15% | 100% |
| Acfer 094 | 3.00 | 0% | 32% | 0% | 0% | 0% | 5% | 0% | 37% | 24% | 7% | 31% | 6% | 2% | 5% | 0% | 19% | 100% |
| Felix | 3.3/3.4 | 0% | 41% | 0% | 0% | 0% | 13% | 7% | 61% | 30% | 0% | 30% | 3% | 2% | 4% | 0% | 0% | 100% |

Type – Petrologic Type; Olv – olivine; Fo – forsterite; En – enstatite; ClinEn – clino-enstatite; Mag – magnetite; Ph – phyllosilicate; Am – amorphous matrix.

**3.3 Reflectance Spectra Data Collection**

The reflectance data for all the carbonaceous chondrites measured in this study were acquired at the Brown University NASA/Keck Reflectance Laboratory (Pieters and Hiroi, 2004). Visible/near-infrared measurement were obtained with the bi-directional spectrometer from 0.3-μm to 2.5-μm at a resolution of 10 nm with an incidence angle of 30° and emergence angle of 0°. Pressed halon, with an approximately Lambertian surface, was used as a reflectance standard. Mid-infrared spectra were measured using Thermo Nexus 870 FT-IR spectrometer. This off-axis bi-conical spectrometer uses a brushed gold standard. Data were obtained from 1-μm to 25-μm at a resolution of 4 cm$^{-1}$.

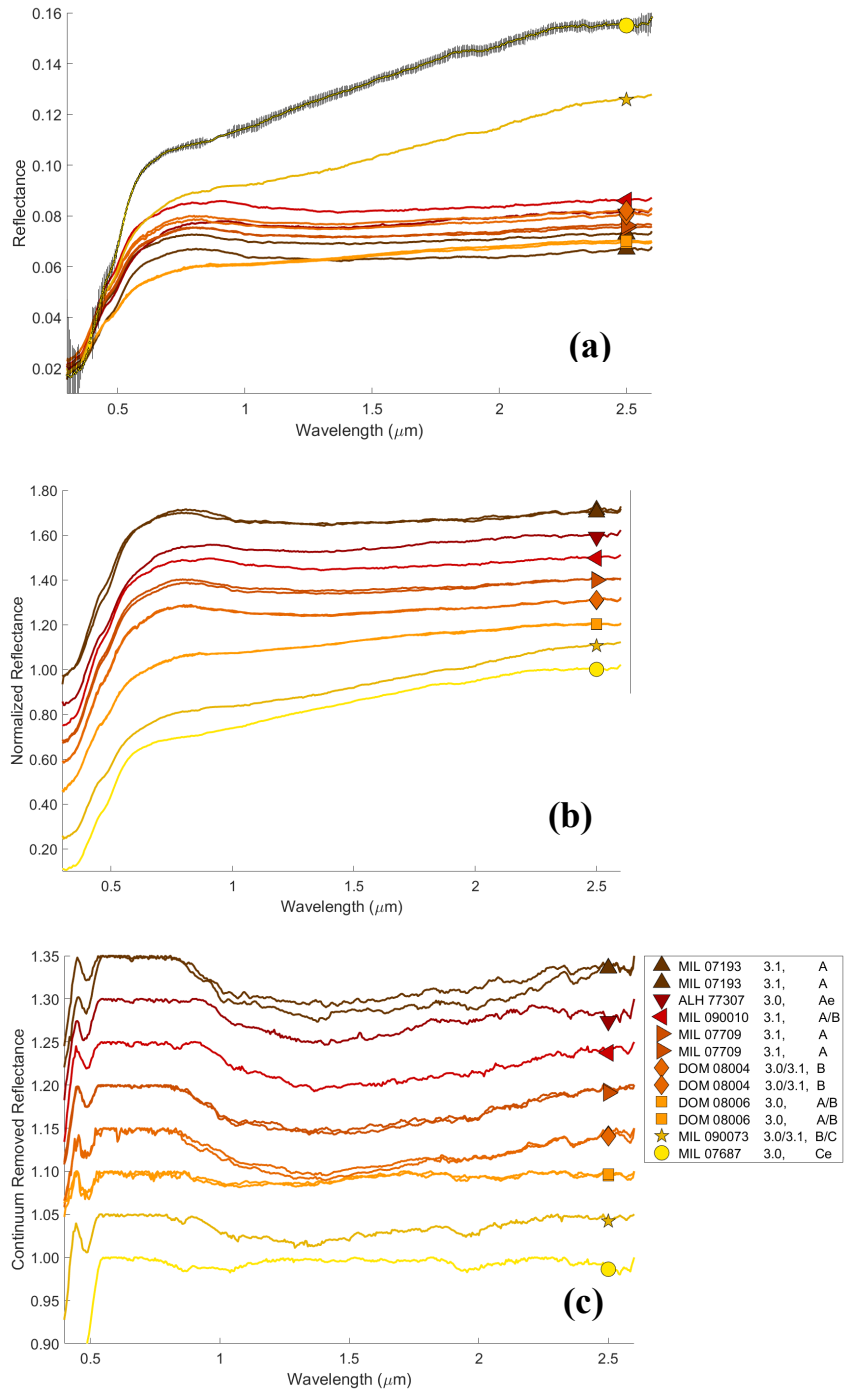

**Figure 2**: (a) Near-infrared data, (b) normalized and offset (+0.5) and (c) continuum removed and offset ordered by weathering grade (top, minimally weathered to bottom, most weathered). Error bars are shown for MIL 07687 in (a). The color and symbol identify each sample. For minimally terrestrially weathered low-metamorphic grade COs, near-infrared spectra are fairly flat without strong features. Terrestrial weathering increases red slope and causes weak absorptions at 0.9- and 1.9-µm. Upon continuum removal, a broad weak feature (~2-5%) appears in the most pristine samples centered near 1.4-µm. This feature is masked by weathering products in weathered samples. A similar feature has previously been interpreted as the presence of iron-bearing silicate matrix (Cloutis, et al., 2012).

## 4. Results

### 4.1 Near-infrared Results

The visible/near-infrared reflectance spectra of the suite of eight least-processed CO samples are presented in **Figure 2**. A convex hull from 0.3-2.5-μm was used to remove the overall slope in the spectra, and enhance any features present. A broad weak feature centered near 1.4-μm is apparent for meteorites that have minimal terrestrial weathering. This feature has been previously attributed to amorphous iron-bearing silicates (Cloutis et al., 2012).

Weathering of the meteorites affects the near-infrared spectral slope, causing significant reddening (**Figure 3**). A 4$^{th}$-order polynomial was fit to the data using standard curve fitting routines in MATLAB and the minimum for each spectrum is then determined. For the minimally weathered samples the position of the band is found to be at ~1.4-μm (**Figure 4**). With increasing weathering grade, the position of the band shifts to shorter wavelengths, until it is not observable (e.g., MIL 07687). Weathering also reduces the depth of this feature (**Figure 4**). These changes are caused by hydrated iron oxides and gypsum produced during terrestrial weathering that mask the 1.4-μm feature. Small features at 0.9-μm and 1.9-μm also appear with significant terrestrial weathering. These are produced by evaporite minerals and terrestrial hydroxides (Velbel et al, 1991; Clark, 1999).

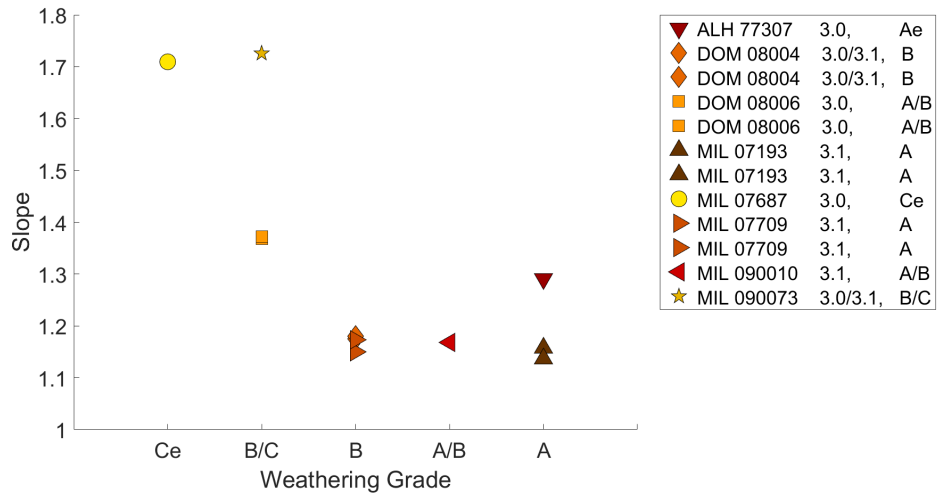

**Figure 3:** Near-infrared slope vs. weathering grade. Slope is defined as the ratio of reflectance values at 0.55- and 2.4-μm. Minimally weathered low-metamorphic grade COs, with weathering grades A, A/B, or B, have a fairly flat slope (~1.1-1.2). Meteorites with more severe terrestrial weathering, grades B/C or Ce, have higher slope values indicating increased reddening.

The 1.4-μm feature in the low-metamorphic grade CO meteorites is qualitatively similar to features seen for certain natural and synthetic glasses (Cannon et al, 2017; Horgan, et al., 2014; Moroz et al., 2009). Generally, tektites, obsidians, and other natural glasses exhibit a weak feature centered near 1.1-1.2-μm (Horgan et al., 2014). Fast-cooled synthetic glasses formed by rapid melting of JSC Mars-1 (an altered volcanic ash analogous to the Martian surface) show features that are similar in strength but at a shorter wavelength than the least-processed CO meteorites (~1.2-μm for fast cooled synthetic glasses compared to 1.4-μm feature of the least-processed COs). JSC Mars-1 has an iron content of 12-15 wt.% (Moroz et al., 2009). Other studies (Cannon et al., 2017) have synthesized glasses with iron contents of ≤25 wt.% to simulate expected impact/volcanic glass compositions for Mars, the Moon, and Mercury. The iron content of synthetic glasses is lower than what is observed in the matrices of least-processed CO meteorites, which have significantly higher iron contents (30-35 wt.% iron; Brearley, 1993). The near-infrared differences in wavelength between the least-processed COs and natural/synthetic glasses may be attributable to the iron content of the CO meteorite samples. In addition, since the amorphous matrix materials in these COs are olivine-normative, it is possible that the near-infrared spectral behavior will be similar to iron-rich olivines (Dyar et al., 1998). Fayalites have a ~1-μm complex feature caused by three spin-allowed crystal field transitions of irons in the M1 and M2 sites (e.g., Burns and Huggins, 1972; Dyar et al., 2009). We therefore suggest that since these amorphous silicates have olivine-normative compositions, the near-infrared spectral features for the pristine (minimally terrestrial weathered) least-processed CO meteorites are caused by the amorphous iron-bearing silicate matrix materials. However more experiments are required to determine the exact cause of this feature.

**Figure 4:** Band position vs. band depth. Terrestrially weathered samples have weaker (smaller band depths) that are centered ~1.2-μm. Minimally weathered low-metamorphic grade COs have bands around 1.4-μm, with band depths of 5-7%. The weathered samples have band with positions at shorter wavelengths and band depths of 0-2%. Color indicates the weathering grade, dark red symbols are less terrestrially weathered while lighter, yellow colors have more significant weathering.

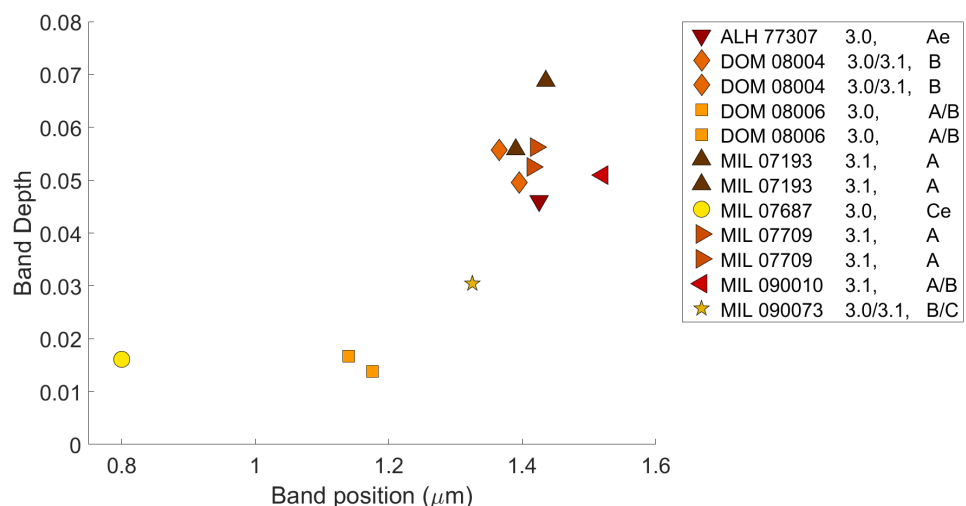

### 4.2 Mid-infrared Results

In the mid-infrared (**Figure 5**), least-processed CO meteorites exhibit features consistent with olivine and pyroxene in the chondrules, with strong absorption features centered at ~12.7-, 19.5- and 24-μm. Additional overlapping absorptions on the short-wavelength side of the 12.7-μm feature are controlled by relative abundance of olivine and pyroxene. Some meteorites, for which two samples are measured, differ slightly primarily in slope (especially at long wavelengths). However, the positions of the features are the same. This is likely caused by differences in how the sample is placed in the sample dish (Salisbury and Eastes, 1985; Salisbury and Wald, 1992). Compared to the pristine least-processed COs, terrestrially weathered samples have minor spectral differences in the mid-infrared wavelengths. MIL 090073 and MIL 07687 have a more extreme slope between 10- and 11.5-μm. However, the spectral features of olivine (12.7-, 19.5-, and 24-μm) are still apparent despite the weathering.

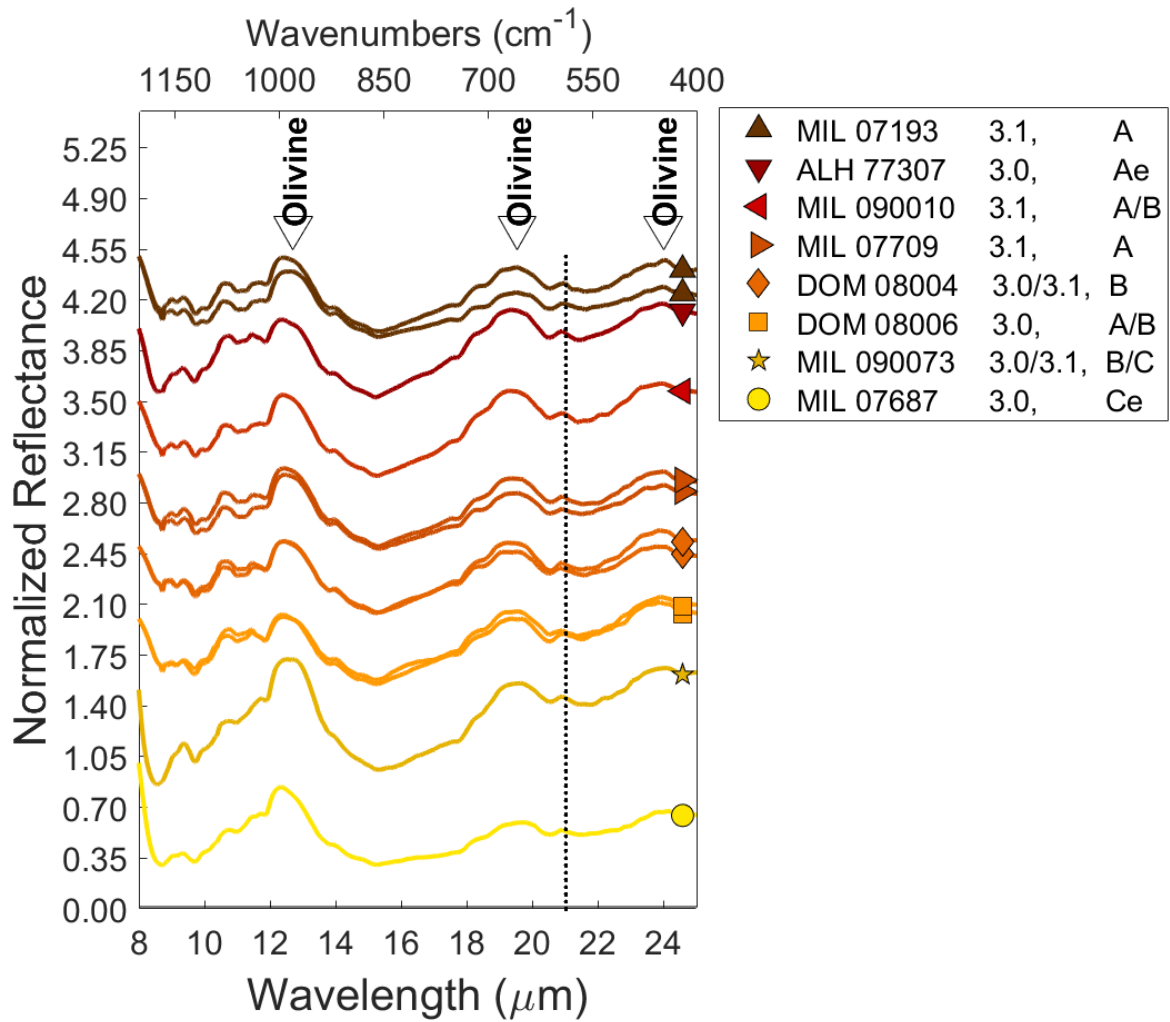

**Figure 5:** Mid-infrared spectra of low-metamorphic grade CO meteorites. Normalized, offset data for eight CO meteorites ordered top to bottom by weathering grade (minimally terrestrially weathered samples at the top, more severely weathered samples at the bottom). Low-metamorphic grade CO meteorites exhibit strong features caused by olivine and pyroxenes in chondrules with features at ~12.7, 19.5- and 24-μm. An additional feature at ~21-μm is present in each spectrum. This is the Si-O vibration in the amorphous silicate matrix. Dashed lines indicate the positions of Si-O vibration in amorphous silicates and glasses.

In all of the spectra of the least-processed COs, a feature can be seen at 21-μm that is attributable to Si-O bending vibrations in amorphous iron-bearing silicate matrix. This feature is ubiquitous and characteristic of least-processed CO meteorites in this wavelength region. Olivine may have a weak feature near 21-μm (e.g., Koike et al., 2003; Hamilton, 2010 and references therein). The position of this feature changes depending on the composition of the olivine. The feature in fayalitic olivines occur at a similar position to the 21-μm feature observed in the least-processed COs. However, fayalitic olivine in the least-processed COs is rare (e.g., Davidson et al., 2016; DeGregoio et al., 2016; Alexander et al., 2018) so any contribution to this feature from fayalite is likely minimal. Crystalline silicates, mostly in chondrules, are dominated by magnesium rich olivine (Fo70 or greater). In forsteritic olivine, the absorption is at 21.7-μm (e.g., Hamilton, 2010; Koike et al., 2003) at slightly longer wavelengths than 21 μm feature present in the spectra of CO meteorites. Thus olivine is not contributing significantly to the 21-μm feature characteristic of low-metamorphic CO meteorites. This 21 μm features is however consistent with amorphous silicates.

To further examine the contribution of the amorphous matrix to the mid-infrared spectra, the crystalline component of each spectrum was approximated with a simple linear model (Thompson and Salisbury, 1993; Ramsey and Christensen, 1998) with four components (**Table 3**) including two olivines with different compositions (~Fo90 and Fo11; Sunshine and Pieters, 1998; RELAB database), one enstatite (Bamble, Norway; $Mg_{.88}Fe_{0.14}SiO_3$; RELAB database), and a neutral darkening agent. The standards used as model components have <45-μm grain size. These spectral endmembers, like the spectra of CO meteorites, were also collected at RELAB. The olivines were chosen to have features near 21 μm to ensure that this feature is included in the modeling of these meteorites. These components represent the approximate range of mafic minerals found in the least-processed CO chondrules and lithic fragments (e.g., Davidson et al., 2014; **Table 2**). The neutral darkening agent is a constant that is used to lower the spectral contrast of the olivine and pyroxene endmembers used in the model to improve the fit between the model and data. The component spectra are presented in **Figure 6**. As in previous spectral modeling efforts, these models rely on linear mixing of spectral component endmembers (e.g., Thompson and Salisbury, 1993; Ramsey and Christensen, 1998). Other mid-infrared spectral models use large databases of minerals to determine the relative abundances of minerals in a given spectrum. The spectral models presented here are a first order removal of the crystalline components of the meteorite spectra. The best fit models were obtained by using the non-negative least-squares fitting routine in MATLAB that uses matrix decomposition to determine the relative fractions of each spectral endmembers in a given meteorite spectra. The mid-infrared spectral models CO chondrites analyzed here are not intended to estimate modal mineralogies, but instead to broadly remove the crystalline components in the meteorite spectra in order to emphasize the spectral features of the amorphous iron-bearing matrix. An example model fit to ALH 77307 is shown in **Figure 7,** along with the residual spectrum.

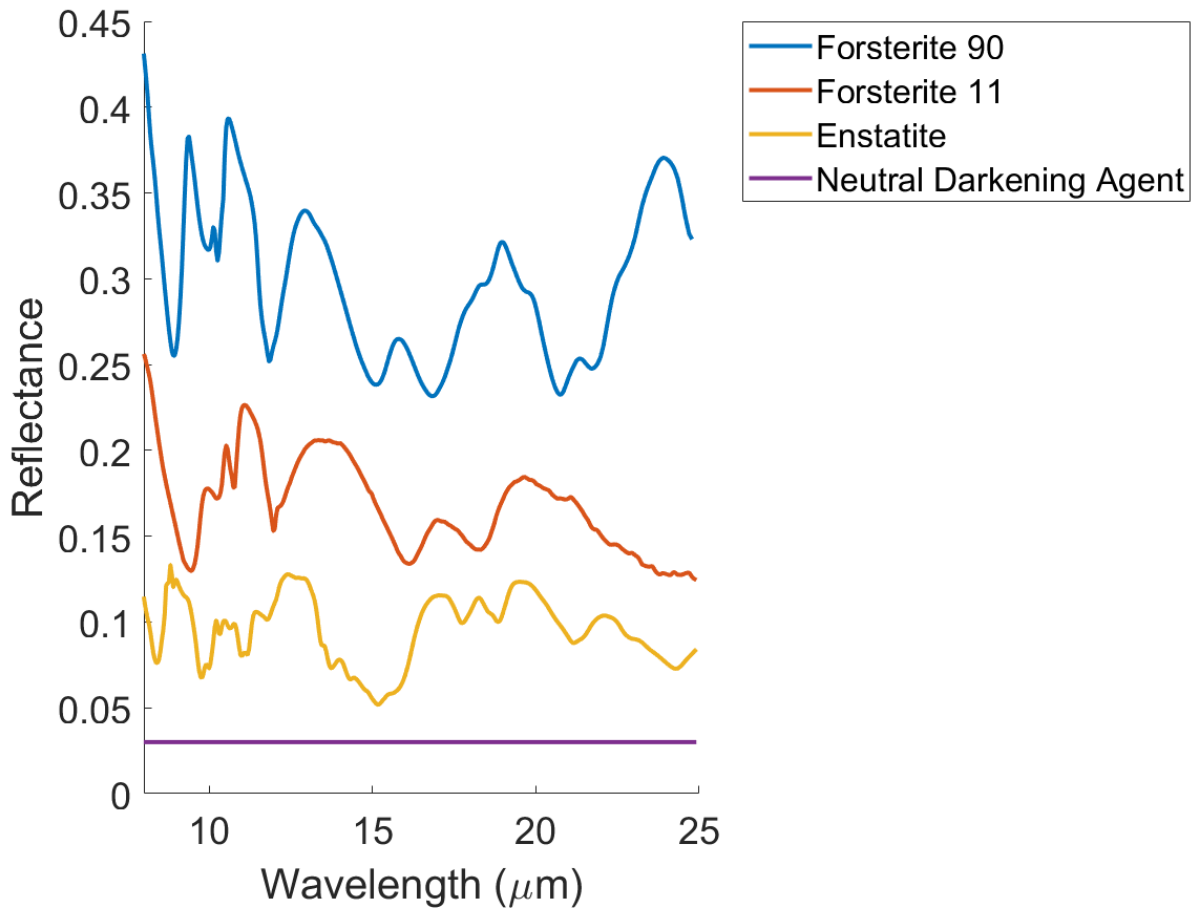

**Figure 6:** Four component spectra used for modeling (offset): two olivines of differing iron content (~Fo90 and Fo11), one enstatite, and a neutral darkening agent. These are used as a first order model of the olivine and pyroxene components to enhance the spectral signature of the amorphous iron-bearing silicates characteristic of the matrices of low-metamorphic grade meteorites.

**Table 3: Spectral Modeling Components**

| Name | Mineral | Grainsize | Origin | References |
|---|---|---|---|---|
| Forsterite 90 (approx.) | Mg-rich olivine | <45-µm | San Carlos Indian Reservation | RELAB Database. Sample ID: BE-JFM-041. |
| Forsterite 11 | Fe-rich olivine | <45-µm | Kiglipait intrusion. | Sunshine and Pieters, 1998; King and Ridley, 1987 |
| Enstatite $(Mg_{0.88}Fe_{0.14})SiO_3$ | Mg-rich pyroxene | <45-µm | Bamble, Norway | Sunshine and Pieters, 1998 |

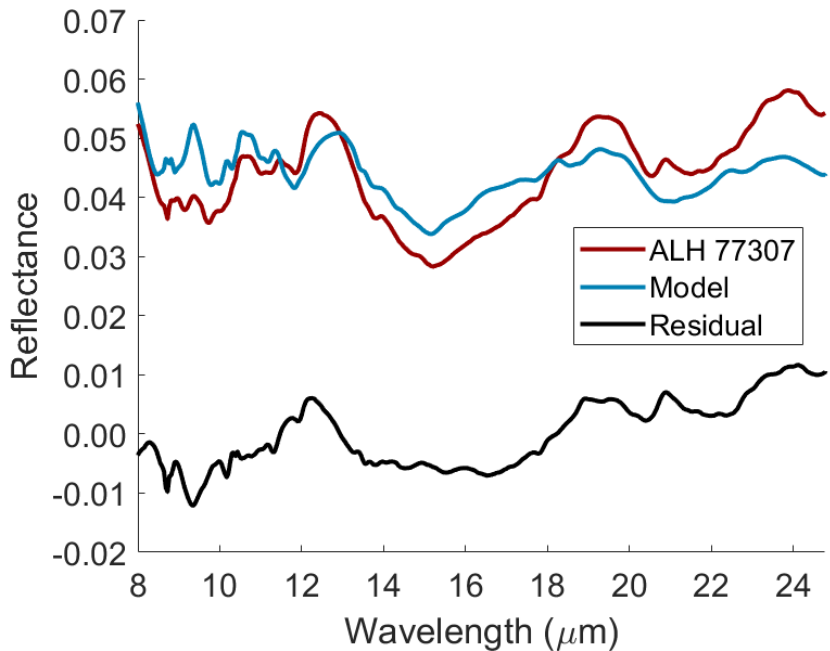

**Figure 7:** Example of mid-infrared spectral modeling. The spectrum of ALH 77307 (red) is modeled using the non-negative linear least-squares approach with the four component spectra shown in Figure 6. The residual is the difference between the data (red) and the model (blue), shown relative to zero (bottom, black). Although the fit is poor in the region between 8 and 12 μm, the amorphous silicate feature 21 μm is enhanced in the residual spectrum. The discrepancy between the model and data is due to the imperfect terrestrial analogs and non-linear mixing effects for the grain sizes of meteorite sample and modeling components.

The residual spectra for all least-processed CO meteorites are presented in **Figure 8**. These residuals have two common characteristics: (1) all have negative residuals in the 8-11-μm region, and (2) all residuals exhibit positive features at ~9.1-, 12.2-, and 21-μm. The negative residuals in the 8-11-μm region are caused by the model mineral spectra having significantly stronger features than the COs in this region. This may be caused by non-linear mixing in this spectral region, which can occur for samples with small grain sizes (e.g., <75-μm, Hunt and Vincent, 1968; Hunt and Logan, 1972; Salisbury et al., 1992; Salisbury and Eastes, 1985). Additionally, interactions between low albedo materials, such as the amorphous silicates (approximated by the neutral darkening agent), and higher albedo crystalline silicates are poorly understood. While some features in the residual spectra may be caused by limitations in endmembers and the inadequacies of linear mixing models for fine particulates, the features in the residuals at 9.1, 12.2 and 21-μm are qualitatively similar to the spectrum of condensed iron-magnesium-smokes (Fe smokes; Hallenbeck et al, 2000; Hallenbeck et al., 1998; **Figure 8**). An example of the spectrum of a condensed Fe-smoke is shown in **Figure 8** (Hallenbeck et al, 1998). Fe-smokes were created in the laboratory to simulate amorphous phases inferred in the coma of comets and the interstellar medium from mid-infrared spectra (e.g., Hallenbeck et al., 1998, 2000). In samples from comet Wild 2 returned by the Stardust Mission, amorphous phases appear to be similar to GEMs grains found in chondritic-porous inter-planetary dust particles (e.g., Ishii et al., 2008). The amorphous materials characteristic of the least-processed COs are similar to GEMs material in texture and composition (Ishii et al., 2008). Therefore, the smokes presented by Hallenbeck et al. (1998, 2000) represent the best available analog materials for the amorphous phases found in least-processed CO meteorites.

In particular, the spectra of the smokes are qualitatively similar to the residuals from modeling CO meteorites with specific features at ~9.1-µm, a weak feature at 12.2-µm and a stronger 21-µm. The differences between the meteorite residuals and the Fe-smokes are the relative strengths of the vibrations and the slope at long-wavelengths. Since the matrix materials in the least-processed COs do not have the same compositions as the smokes, the vibration strengths and positions are expected to be slightly different than the example presented. Additionally, the spectral model for the crystalline components is a first-order approach and, as a result, the residual spectra almost certainly contain artifacts that contribute to the spectral discrepancies. The features at 9.1- and 12.2-µm need to be investigated further before they can be uniquely attributed to amorphous phases.

Si-O bending vibrations are known to produce features at 21-um in the spectra of amorphous Fe-bearing silicates (Mysen, 1982; Mysen et al., 1982; McMillan, 1984a, b, Dorschner et al., 1995). Since crystalline fayalite and forsterite are not likely to contribute significantly to the 21-µm feature in these samples, the 21-µm feature can be confidently attributed to Si-O bending vibrations in the amorphous iron-bearing silicates that are abundant in the studied least-processed CO meteorites (e.g., Howard et al., 2015; Alexander et al., 2018).

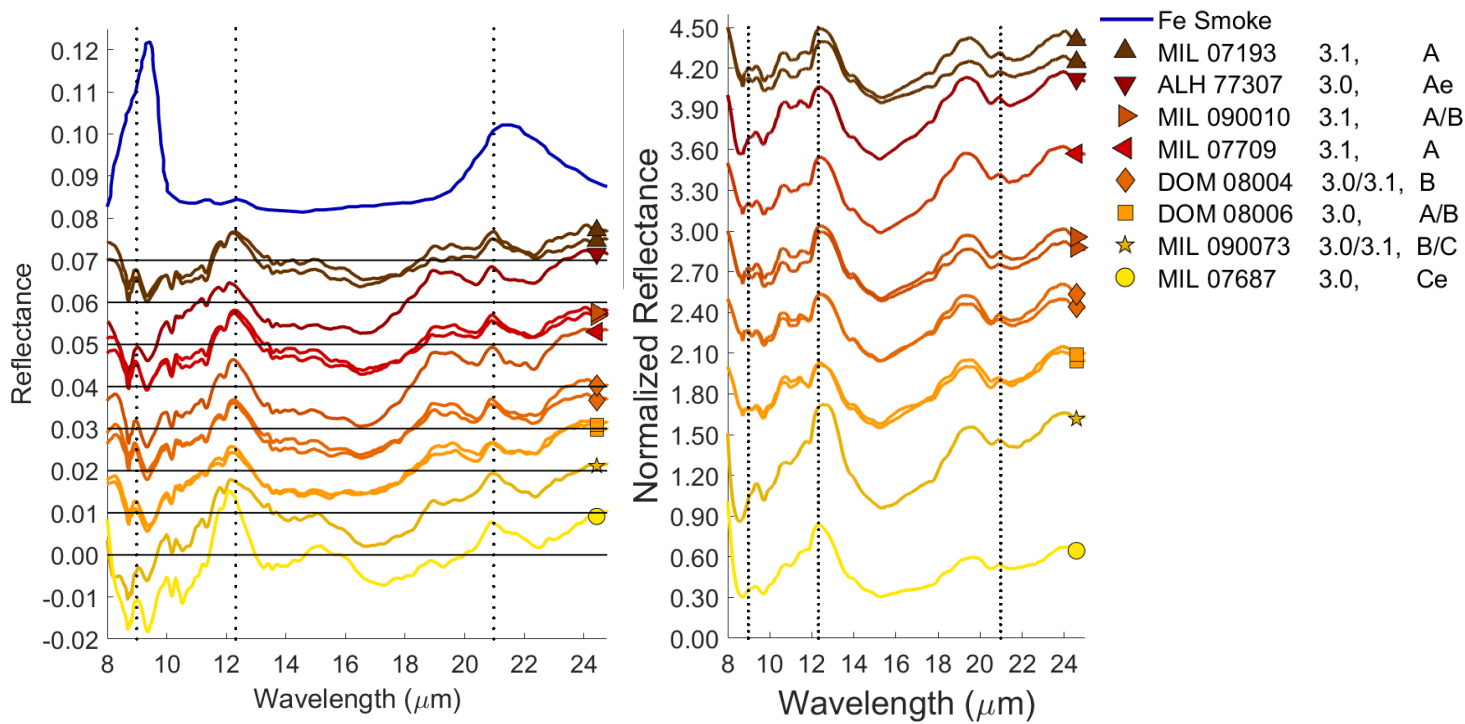

**Figure 8:** Residual spectra (after crystalline models are subtracted from data) for each of the low metamorphic grade CO meteorites. The positions of zero are indicated for each residual and are offset for clarity. They are ordered by weathering grade (pristine samples at the top). The residuals are compared to the mid-infrared spectrum of an Fe-smoke. The Fe-smoke is a laboratory condensate made of iron, silicon and oxygen (Hallenbeck et al, 1998) that is the best analog for amorphous phases found in the least-processed meteorites. The residuals are qualitatively similar to the Fe-smoke indicating that the 21-μm feature, in particular, is likely caused by the amorphous iron-bearing phases characteristic of these meteorites. The qualitatively similar features at 9.1-, 12.2 and 21-μm are indicated by the dotted lines. These first approximation models for the crystalline components poorly reproduce the 8-12-μm region where the residuals are consistently negative. This is likely caused by non-linear mixing of fine particulates in this region. Differences between the Fe-smoke and the meteorite residuals are the relative strengths and positions of the Si-O vibrations. These differences are likely caused by the imperfect modeling techniques used and compositional differences between the Fe-smoke and the amorphous iron-bearing matrix silicates.

The feature near 21-μm is the first mid-infrared spectral identification of the amorphous iron-bearing matrix material characteristic of least-processed carbonaceous chondrites.

### 4.3 Comparison to metamorphosed COs

The secondary modification of the CO chondrites is dominated by the effects by thermal metamorphism in their parent body. While the least-processed COs have experienced minimal processing, they appear to have been heated to ~200-300ºC. The more thermally metamorphosed COs experienced recrystallization of the amorphous matrix, followed by coarsening of the matrix, progressive equilibration between matrix and chondrule minerals, and recrystallization of the chondrule mesostasis. The maximum metamorphic temperatures experienced by any known members of the CO chemical group were ~700ºC (McSween, 1977; Cody et al., 2008). These temperatures are high enough to recrystallize all amorphous material as well as equilibrate matrix and chondrule olivine compositions.

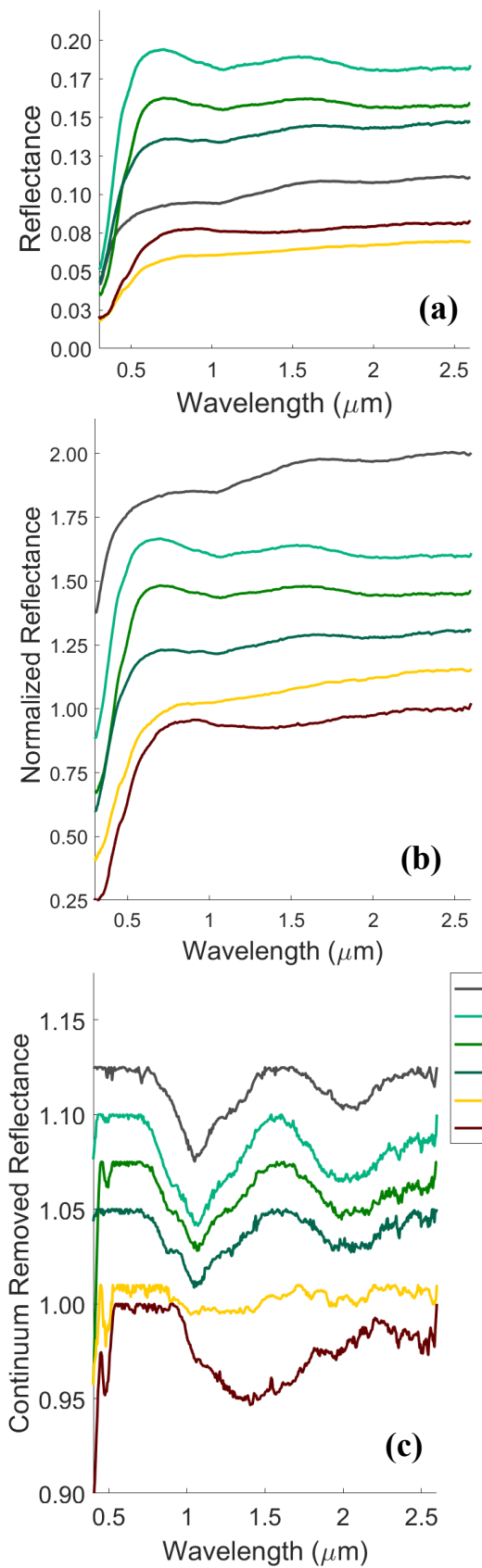

**Figure 9:** Near-infrared comparison to metamorphosed COs. Low-metamorphic grade COs, ALH 77307 and DOM 08006, are compared to three metamorphosed COs: Felix (3.3/4), ALH 85008 (3.5) and ALH 83108 (3.5), in green and Allende (CV3.6, gray). (a) reflectance, (b) normalized offset (+.15) and (c) continuum removed reflectance. During metamorphism, minerals equilibrate creating more crystalline olivine with homogenous composition. Features in the near-infrared for metamorphosed samples reflect the olivine, pyroxene composition with features at 1- and 2-μm. Metamorphosed COs are more similar to CV3 meteorites (e.g., Allende), which are also metamorphosed, than to the relatively less-processed meteorites in their own chemical group.

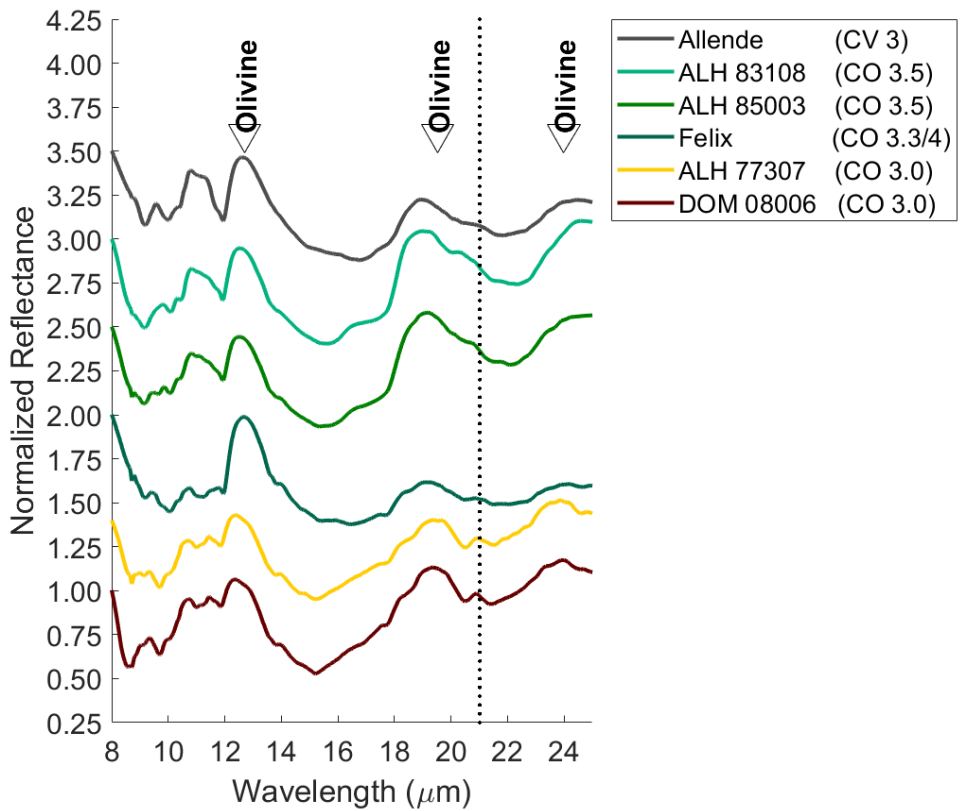

**Figure 10:** Mid-infrared comparison to metamorphosed COs. Spectra of two pristine, low-metamorphic grade COs are compared to thermally metamorphosed COs, ALH 83108 and ALH 85003, and one CV, Allende. After reaching metamorphic grade of 3.3/3.4, Si-O features caused by the amorphous matrix disappear as metamorphism recrystallizes the amorphous materials into crystalline olivines and pyroxenes. The most thermally metamorphosed meteorites are more similar to CV meteorites.

Felix is an example of a moderately metamorphosed CO. Its modal mineralogy, derived from PSD-XRD measurements, indicates that Felix has negligible amounts of amorphous material in its matrix and can be modeled using crystalline standards alone (see Section 3.2). Spectrally, this meteorite is significantly different from the least-processed COs, showing much stronger olivine features in the near- and mid-infrared spectral regions (**Figures 9 & 10**). Two higher metamorphic grade COs are also compared to least-processed examples, ALH 85003 (CO 3.5), and ALH 83108 (CO 3.5). These samples were prepared similarly to the least-processed COs and their spectra were also obtained at RELAB. ALH 85003 and ALH 83108 have been heated to higher peak temperatures than Felix and correspondingly exhibit stronger features caused by progressive increase in crystallinity and equilibration of crystalline phases, as well as further recrystallization of remaining chondrule mesostasis, during metamorphism. This has resulted in near-infrared features at 1- and 2-μm caused by equilibrated olivines. These features have replaced the 1.4-μm feature of the more primitive material. In the mid-infrared, mineralogical changes affect the spectra by increasing the strength of the ~12.7- 19.5-, and 24-μm features. Additionally, this seems to significantly affect the 8-11-μm region where stronger olivine features appear. Carbonaceous chondrites in other heated groups, such as the CVs, contain similar minerals and have similar spectral features to the thermally metamorphosed CO meteorites. This is also shown in **Figures 9 & 10**, where the CV3.6 Allende is compared to the CO meteorites of various metamorphic grades. All of the meteorites that have experienced more extensive heating than the least-processed CO meteorites do not exhibit a 21-μm feature.

**4.4 Comparison to other 3.0 Carbonaceous Chondrites**

Least-processed meteorites are not limited to the CO chemical group. Several other meteorites have been reported to have amorphous matrices similar to what has been described in the least-processed COs. Specifically, the CRs Meteorite Hills (MET) 00426 and Queen Alexandra Range (QUE) 99177 (Abreu and Brearley, 2010; Le Guillou and Brearley, 2014) and the ungrouped chondrite Acfer 094 (Greshake, 1997).

The CR chemical group of carbonaceous chondrites is characterized by aqueous alteration. The CR3 meteorites may have experienced some interactions with water, however the amorphous phases are preserved in the CR3 samples (e.g., Abreu and Brearley, 2010; Le Guillou and Brearley, 2014). These phases are preserved because conditions in the region of the parent body(s) prevented the formation of phyllosilicates since the interaction with water was minimal in these parent bodies/regions of parent bodies (Abreu and Brearley, 2010). Amorphous materials formed through disequilibrium condensation are also preserved in CM meteorites Paris and Y-791198 (Leroux et al., 2015; Chizmadia and Brearley, 2008). Unlike the CR3s, Paris and Y-791198 have evidence for extensive, but heterogeneous aqueous alteration (e.g. the presence of large abundances of phyllosilicates; e.g., Howard et al., 2011; Marrocchi et al., 2014). This heterogeneity allowed some of the initially accreted amorphous materials to be preserved. In contrast to these CMs, the CR3 meteorites, which do not show such extensive evidence for aqueous alteration, represent a parent body or region of a parent body that has undergone little change since the time of accretion. For this study, we confine ourselves to comparisons between the low metamorphic grade COs and these relatively least-processed CR3 meteorites and Acfer 094. The CR3s and Acfer 094 are similar to the least-processed COs in terms of abundances of amorphous iron-bearing matrix silicates.

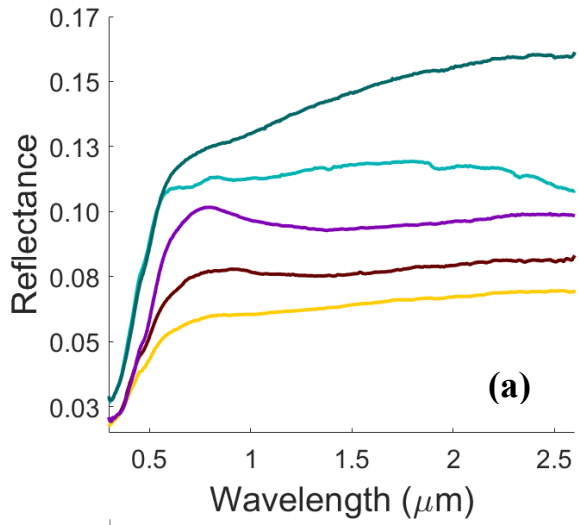
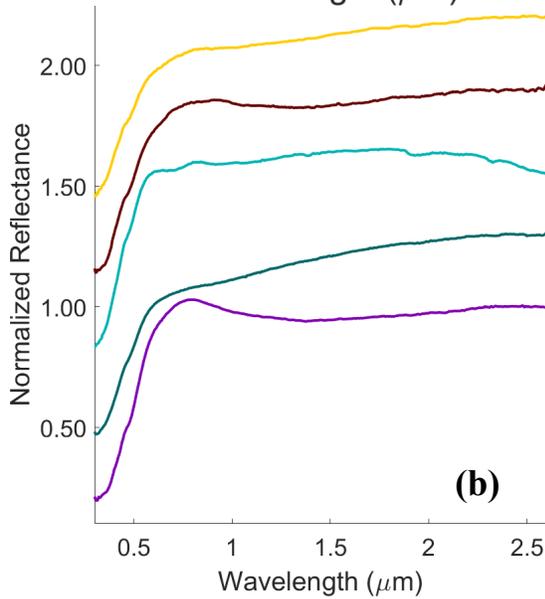
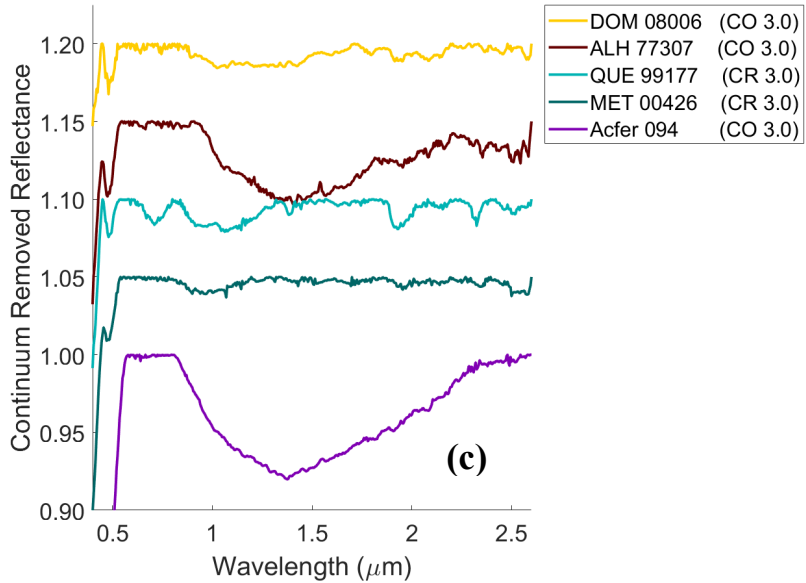

**Figure 11:** Near-infrared comparison to least-processed meteorites in other chemical groups. Near-infrared spectra (a), offset (+0.3) and normalized (b) and continuum removed (c) show that least-processed meteorites of all chemical groups are characterized by low albedos, and where weathering is minimal, they are also characterized by a broad 1.4-µm feature. Terrestrial weathering affects the CRs significantly, with clay and evaporite minerals producing features at 0.7-, 1.0-, 1.9- and 2.4-µm. Acfer 094 has the strongest 1.4-µm of all the low-metamorphic grade carbonaceous chondrites studied. This sample is has significant terrestrial weathering which does not affect the spectrum in the same way Antarctic terrestrial weathering affects the CO3.0 and CR3s. Further investigation is required to determine why Acfer 094's feature is so strong relative to the other 3.0 carbonaceous chondrites.

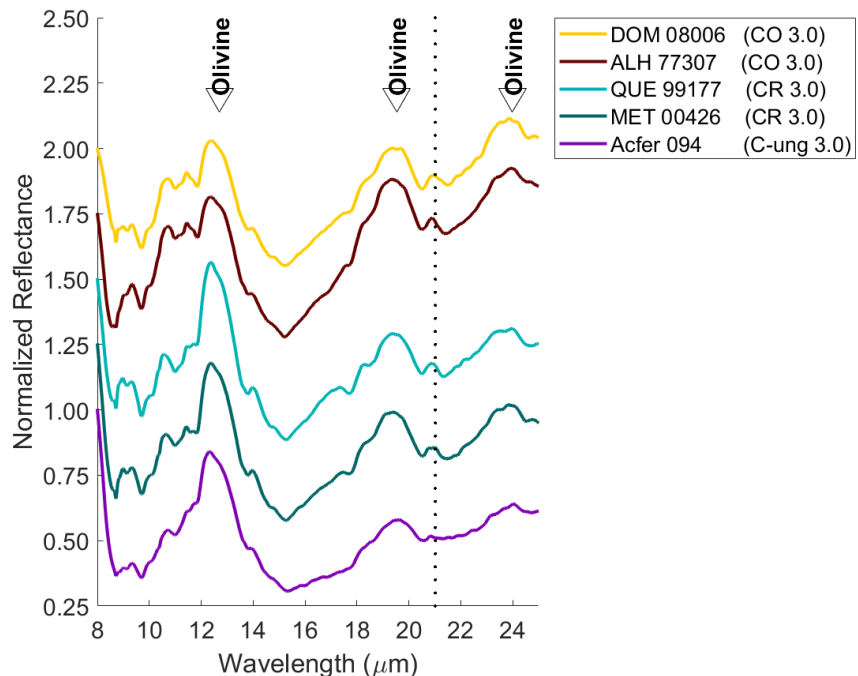

**Figure 12:** Mid-infrared comparison of the spectra of least-processed meteorites in other chemical groups. In the mid-infrared, the spectra of all least-processed carbonaceous chondrites exhibit 12.7-, 19.5- and 24-μm features caused by the olivine and pyroxenes in their chondrules. Additionally, they all have the 21-μm produced by the Si-O bonds from their amorphous iron-bearing silicate matrices.

The CR samples have some terrestrial weathering and thus their near-infrared spectra are comparable to the weathered COs. The CR group of meteorites is generally characterized by larger abundances of metal (e.g., Kallemeyn et al., 1994). Consequently, terrestrial weathering may affect low metamorphic grade CRs differently than COs, particularly the proportions and compositions of terrestrial weathering products. Weathering on the low metamorphic grade CRs, similar to the low metamorphic grade COs, masks the near-infrared signature of the abundant amorphous iron-bearing material. Acfer 094 has a feature at ~1.4-μm that is even deeper than what is observed in the least-processed COs (**Figure 11**). However, this meteorite was found in the Sahara and is likely to be more weathered than the Antarctic COs and CRs presented here. Weathering effects have been compared for 'hot' deserts, like the Sahara, and 'cold' deserts, like Antarctic, finding that the weathering products are generally similar in (e.g., Lee and Bland, 2004). Further investigation is required to understand why Acfer 094 does not have terrestrial alteration features similar to the CRs and the more weathered Antarctic CO meteorites presented here. In the mid-infrared, the CRs and Acfer 094 exhibit similar features to the least-processed COs, particularly at 21 μm, that we attribute to the Si-O vibrations in the amorphous iron-bearing silicates characteristic in these least-processed samples (**Figure 12**).

## 5. Evidence for Amorphous Material on Asteroids

There is evidence for the presence of amorphous materials on at least one asteroid: (93) Minerva. The near-infrared spectrum of Minerva collected with SPeX at the NASA's Infrared Telescope Facility (IRTF) is shown in **Figure 13**. SPeX is a medium resolution cryogenic instrument (Rayner et al., 2003), which has been used to observe a large number of asteroids in the near-infrared (e.g., DeMeo et al., 2009; Clark et al, 2010). The SPeX data are scaled to the visible data from the Small Main-Belt Asteroid Spectroscopic Survey (Xu et al., 1995; Bus and Binzel, 2002; normalized at 0.55-μm) to produce the final spectrum. Observational circumstances and the standard stars used during those observations are listed in **Table 4.** The spectrum of Minerva shown in **Figure 13** is normalized to unity at 2.4-μm and compared DOM 08004 (CO 3.0/3.1; normalized to unity at 2.4-μm; Minerva data is offset by +0.75). Spectrally classified as a C-type (Xu et al., 1995; Bus and Binzel, 2002), Minerva's overall low albedo (0.062 +/- 0.015; Marchis et al., 2013) and broad ~1.4-μm feature are very similar to least-processed COs. Although the spectra of some asteroids, which are thought to be related to angrite meteorites (clinopyroxene dominated achondritic meteorites; Burbine et al., 2006), have some similarities to Minerva and the least-processed COs, they have much stronger features (~15%) and much higher albedos (~8-16%). Minerva has not been observed using mid-infrared spectroscopy. However, based on the presence and similarity of the 1.4-μm feature and its low albedo, Minerva is interpreted to be compositionally similar to the least-processed meteorites.

**Table 4: Minerva Observations**

| Asteroid | Date (UT) | Start Time (UT) | End Time (UT) | Int. (Min) | Airmass | Standard |
|---|---|---|---|---|---|---|
| (93) Minerva | 2003-04-27 | 08:31:05 | 08:59:29 | 18.0 | 1.132-1.137 | Landolt (SA) 107-684, Landolt (SA) 107-998 |

Minerva is a triple system, with two satellites that likely formed as a result of a recent collision (Marchis et al., 2013; Yang et al., 2016). Detailed observations of this system constrain the physical properties of Minerva (e.g., Marchis et al., 2013; Yang et al., 2012). These are presented in **Table 5**. The satellites appear spectrally similar to the primary in the visible and near-infrared (Yang et al., 2016). However, these measurements are difficult to make and heavily affected by the image processing to remove the signal of the primary. Furthermore, these observations have significant atmospheric absorption between 1.3-1.5-μm so that the 1.4-μm feature cannot be detected.

**Table 5: Physical Properties of Minerva System**

| Asteroid | Albedo | Diameter | Mass | Bulk Density |
|---|---|---|---|---|
| (93) Minerva[1] | 0.062 +/- 0.013 | 154 +/- 6 km | 3.35 +/- 0.54 x $10^{18}$ kg | 1.75 g/cm$^3$ |
| (I) Aegis[2] | 0.062 +/- 0.013* | 4.3 +/- 1.5 km | - | - |
| (II) Gorgoneion[2] | 0.062 +/- 0.013* | 3.6 +/- 1.0 km | - | - |

*assumed from Marchis et al., 2013.
[1]Marchis et al., 2013
[2]Yang et al., 2016.

The impact that created the satellites likely occurred relatively recently, ~1 billion years ago, based on estimates on the rigidity of the primary and the specific tidal dissipation (Marchis et al., 2013; Yang et al., 2016). However, this system may be older if it experienced non-tidal effects such as gravitational interactions with other asteroids or late-inward migration of Jupiter (Marchis et al., 2013). The satellite-creating impact likely caused Minerva to have significant macro-porosity (Marchis et al., 2013) that lowers the bulk-density of the asteroid. While this impact was probably large enough to resurface part or all of the asteroid, removing any effects of surface processing (e.g., Thomas and Robinson, 2005), C-type asteroids are not strongly affected by optical and near-infrared spectral changes due to space weathering (e.g., Clark et al., 2002; Lantz et al., 2015). Furthermore, our observations of Minerva are hemispherically averaged. The presence of a 1.4-µm band, therefore, implies that Minerva's surface contains a significant amount of amorphous material over a large fraction of the asteroid. Marchis et al., (2013) also report IRTF+SPeX observations of Minerva. These data are similar to our observation in albedo, but the Marchis spectrum does not appear to have a 1.4-µm feature. While Minerva's rotation period it well known (5.98176+/-0.000004 hours). Between the observations of Minerva presented here and those presented by Marchis et al., (2013), 64710 hours elapsed, or 10,817.87 +/- 0.1 rotations periods. This indicates that our observations were taken at a ~87% difference in phase from the Marchis observations. Although differences between our SPeX and that of Marchis et al. (2013) may suggest some rotational heterogeneity, perhaps related to its impact history, more rotational data is needed.

In order to preserve a significant amount of amorphous iron-bearing phases on Minerva, it must not have experienced significant internal heating, which is possible if the asteroid accreted late or the amorphous materials are part of an outer, unprocessed shell unaffected by interior heating. In the late accretion scenario, Minerva would have accreted ~4 Ma after CAI formation (Sugiura and Fujiya. 2014), missing the peak heat flux from the decay of $^{26}$Al. Without significant internal heating, the interior mineralogy would be similar to the surficial mineralogy (e.g., Grimm and McSween, 1989). If Minerva's composition is homogenous, it could not be the parent body of the CO meteorites that have experienced significant thermal metamorphism. The alternative is that Minerva accreted earlier and did experience significant internal heating, with increasing thermal metamorphism with increasing depth (e.g., onion-shell model; Taylor et al., 1987; Huss et al., 2006). In this case, only its outer layers would have experienced minimal heating (200-250ºC; e.g., Bonal et al., 2016) consistent with the preservation of amorphous material. The major impact that created the satellites of Minerva must then have resurfaced the asteroid and created macro-porosity (Marchis et al., 2013), but left these outer layers largely intact. If so, the Marchis spectra, which lack a 1.4-µm absorption and thus has less amorphous material may represent a region of the asteroid that is nearer the impact site contains material from depth.

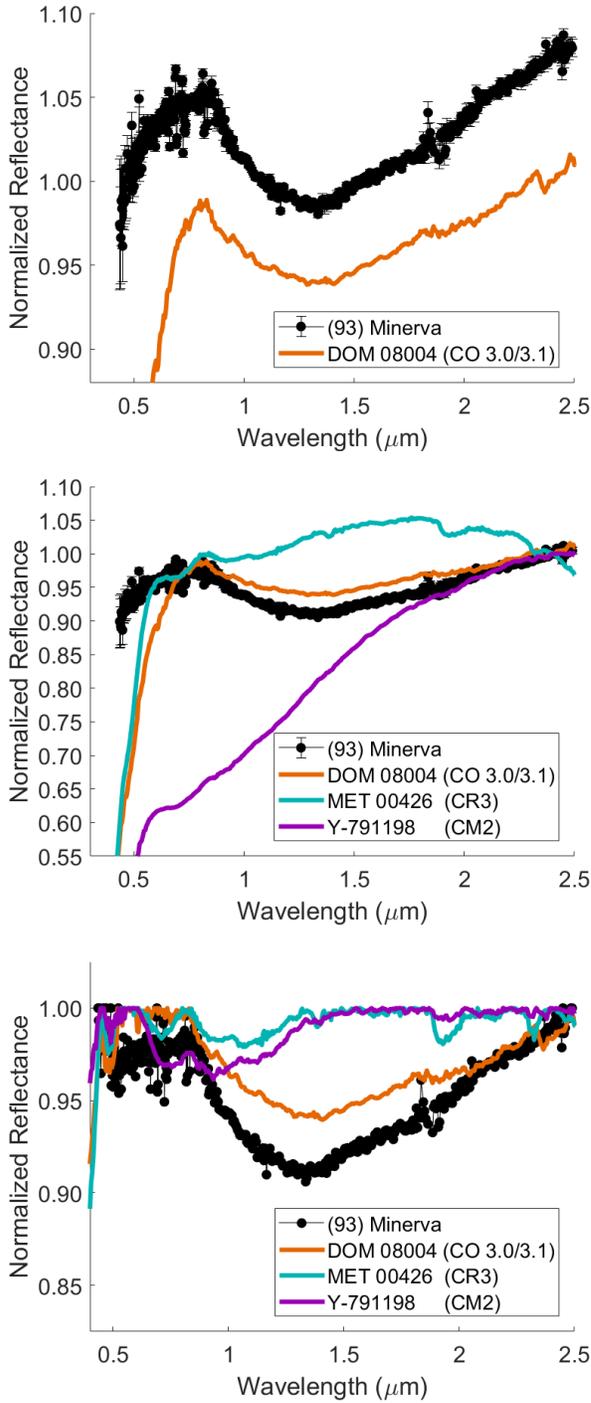

**Figure 13:** (Top) SPeX observation of 93 Minerva (offset +0.75) compared to low-metamorphic grade CO, DOM 08004 (CO 3.0/3.1). (Middle) (93) Minerva is compared to DOM 08004 (CO3.0), MET 00426 (CR3.0) and Yamato-791198 (CM2). (Bottom) continuum removed data are compared. Y-791198 is an aqueously altered meteorite that preserves some amorphous phases at very small scales. MET 00426 has a high abundance of amorphous materials, similar to the low-metamorphic grade CO meteorites. DOM 08004 is most spectrally similar to Minerva. The similarity between the shape and position of the 1.4-μm and comparable low albedo indicates that amorphous materials may be abundant on the surface of Minerva. Minerva is the best candidate for follow up observations in the mid-infrared to confirm the presence of amorphous iron bearing silicates.

It may be possible to distinguish between these scenarios with additional rotationally resolved spectroscopy of Minerva. If the interior materials excavated by recent craters have spectral signatures of thermal metamorphism, the onion-shell model would be confirmed. On the other hand, if there are no rotational heterogeneities, the late-accretion model probably best matches Minerva's history.

While Minerva appears to be related to the least-processed CO meteorites, other asteroids have been suggested to be related to the more extensively thermally metamorphosed CO meteorites. In particular, two K-type asteroids (221) Eos and (653) Berenike have been reported to have near-infrared spectral features similar to those of the thermally metamorphosed CO, Warrenton (CO 3.7; Burbine et al, 2001). Eos and Berenike are dynamically related with semi-major axes near 3.0 AU, but Minerva's semi-major axis is at ~2.7 AU indicating that Minerva is dynamically unrelated to Eos and Berenike. Thus, while compositional evidence is consistent with the CO meteorites coming from a single, internally heated, disrupted parent body (e.g., Weisberg et al., 2006), observational data suggests that least-processed and high metamorphic grade CO-like asteroids both exist in the Asteroid Belt.

Given the similarity between Minerva and the least-processed CO meteorites, it is unlikely that Minerva is related to more thermally metamorphosed meteorites or asteroids Thermally metamorphosed carbonaceous chondrites (e.g., thermally metamorphosed COs and CVs) lose the characteristic 1.4- and 21-µm feature, and have strong olivine features in the mid-infrared (McAdam et al, 2015b). However, Minerva may be related to other meteorites that preserve amorphous phases. Some meteorites that experience the early stages of aqueous alteration may also preserve some amorphous silicates and pre-alteration textures, which has been observed for the CM meteorites Paris and Yamato-791198 (Leroux et al., 2015; Chizmadia and Brearley, 2008). A few ordinary chondrites (e.g., Semarkona and Bishunpur) also have amorphous phases in their matrices (Alexander et al., 1989; Dobrica and Brearley, 2011), but since matrix materials make up much smaller volume fractions in the ordinary chondrites than in CM, CO and CR chondrites, the amorphous material is unlikely to produce strong features in the spectra of these ordinary chondrites. Yamato-791198 and CRs with petrologic grades 1 and 2 (extensively aqueously altered meteorites; e.g., Howard et al., 2015b), are characterized by the presence of large volumes of phyllosilicates produced during aqueous alteration that have strong mid-infrared spectral features consistent with phyllosilicates (McAdam et al., 2015a, b). While Minerva does not have a 0.7-µm absorption, indicative of the presence of phyllosilicates, mid-infrared observations are required to determine if phyllosilicates are present (e.g., McAdam et al., 2015a). The CRs presented in **Figure 13** have experienced some terrestrial weathering, similar to the weathered COs presented in **Figure 2**. It is therefore unclear if Minerva is related to the least-processed meteorites in the CO or the CR chemical group. There are subtle differences between the least-processed CRs and COs (e.g., **Figure 12**) in the mid-infrared (e.g., the slope in and relative strength of features in the 8-12-µm region). Additionally, terrestrial weathering does not affect the mid-infrared range as significantly as the near-infrared so mid-infrared observations may be useful for determining which chemical group Minerva is related to.

The target asteroids of the on-going sample return missions OSIRIS-REx and Hayabusa-2, (162173) Ryugu and (101955) Bennu, respectively, will soon become the standards of comparison for low-albedo asteroids. Based on the presence of the 0.7-µm feature (Vilas, 2008), Ryugu appears to be an aqueously altered asteroid and so it is unlikely to have significant amorphous materials on its surface. The spectrum of Bennu does not have either the 1.4-µm feature or 21-µm and, therefore, Bennu does not appear to have amorphous-iron bearing silicates on its surface either. However, mid-infrared observations of Bennu did not have sufficient signal-to-noise to detect the 21-μm, Si-O feature (Emery et al., 2014). The vastly improved spatial resolution of the upcoming orbital observations of Bennu will help determine if Bennu contains amorphous silicates.

In addition to the sample return missions to asteroids, the upcoming James Webb Space Telescope (JWST) will have both near- and mid-infrared instruments capable of observing asteroids. The Mid-Infrared Instrument (MIRI) has spectral coverage from 5-28-μm with medium-high resolution. The resolving power ($\lambda/\Delta\lambda$) of this spectral mode is ~1550-3250 (Swinyard et al., 2004; Wells et al., 2015). This resolving power is somewhat high to determine the surface mineralogy of Minerva or other asteroids, leading to long exposure times to achieve the required high signal-to-noise ratios. However, binning the data to resolution of ~500 or less should allow the identification of mineralogy including the presence of the 21-μm feature indicative of amorphous iron-bearing silicates. The signal-to-noise ratio (SNR) requirements for observing the diagnostic feature of amorphous materials at 21-μm are estimated by resampling the laboratory spectra of CO meteorites to a spectral resolution ~500, then adding noise to that spectrum and determining empirically the minimum SNR where the 21-μm feature is detectable. We find that a SNR of ~50 could allow a marginal detection (e.g., **Figure 14**). A detection can be reliably made, however, with a SNR of ~100.

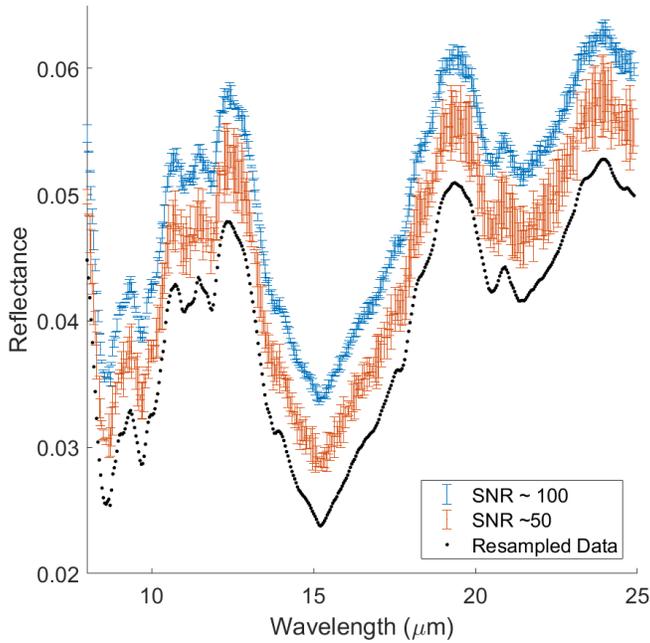

**Figure 14:** SNR test. To estimate the minimum required SNR to detect the 21-μm feature, Gaussian distributed noise is added to laboratory data of ALH 77307 that has been resampled R ~200. We find that a SNR of ~50 could produce a marginal detection, if binning is employed, however, a SNR ~100 or greater is would ensure a detection. We suggest a minimum SNR of ~100 to detect this material on the surface of an asteroid.

## 6. Conclusions

Carbonaceous chondrites that have experienced little thermal processing or aqueous alteration are characterized by amorphous iron-bearing silicates in their interchondrule matrices and chondrule rims. Least-processed meteorites occur in several chemical groups, particularly among CO meteorites, which contain ~30% amorphous materials (e.g., Alexander et al., 2018). We have presented visible/near-infrared and mid-infrared spectroscopy of a suite of 8 least-processed COs. Previous near-infrared spectral studies of the primitive CO ALH 77307 reported the presence of a ~1.3-μm feature that was attributed to the iron-bearing amorphous silicates (Cloutis et al., 2012). We find that this spectral feature, centered at 1.4-μm in our analysis, is characteristic of all least-processed CO meteorites that have had relatively minor terrestrial weathering. Additionally, we present for the first time mid-infrared spectroscopy for least-processed CO meteorites. In the mid-infrared, least-processed COs are characterized by chondrule olivines and pyroxenes. However, an additional feature at 21-μm is present in the spectra of all least-processed COs. This 21-μm feature it attributed to the Si-O vibrations of the amorphous iron-bearing silicate matrix in these samples. This is the first mid-infrared spectral evidence of amorphous material reported for any least-processed carbonaceous chondrite. Upon modeling the crystalline spectral component of these data, additional spectral features at 9.1 and 12.2-μm are revealed. These features are qualitatively similar to those seen in synthetic glasses, however, they may be caused by non-linear mixing of fine particulates. Further study is needed to identify additional spectral features of the amorphous silicates.

We also present the near- and mid-infrared spectra of other least-processed carbonaceous chondrites, the CRs MET 00426 and QUE 99177 and the ungrouped Acfer 094. We find that all least-processed carbonaceous chondrites, regardless of chemical group, exhibit similar spectral features in the near- and mid-infrared; specifically they all exhibit at 1.4- and a characteristic 21-μm features.

Finally, we present visible and near-infrared spectrum of asteroid (93) Minerva. The primary asteroid of this triple system is low in albedo and exhibits a 1.4-μm feature, similar to the least-processed COs. We interpret Minerva to have a fresh surface dominated by amorphous iron-bearing silicates exposed after a relatively recent (~ 1 Ga) impact event. The presence of amorphous iron-rich materials on Minerva's surface could be explained in one of two ways, either: (1) Minerva is an asteroid that accreted late, escaping extensive internal heating from short-lived radioactive nuclei and, therefore, the surface mineralogy is representative of the entire asteroid, or (2) Minerva accreted earlier, was internally heated and has an onion-shell structure with increasing thermal metamorphism with increasing depth. In the latter case, the outer layers of the asteroid experienced less heating than interior layers and the amorphous materials are preserved. Using rotationally resolved spectroscopy, especially in the mid-infrared, it may be possible to distinguish between these scenarios. JWST and the MIRI instrument will be able to help constrain the history of Minerva and similar asteroids related to the low metamorphic grade CO meteorites. A reliable detection of the 21-μm feature can be made with an SNR of ~100 if binning of the 5-28-μm region is employed.

The preservation of amorphous materials on the surface of Minerva and potentially other asteroids provide strong evidence that at least their exteriors have experienced minimal heating and aqueous alteration. Thus, asteroids rich in amorphous materials provide a unique window into the earliest epochs of the Solar System, prior to the onset of parent body processing, and offer an opportunity to explore the compositions and physical processes that occurred in the nebula and the early accretionary phases of asteroid formation.

**Acknowledgements:**

Spectra were acquired using the NASA Keck RELAB, a multiuser facility at Brown University. The efforts of Dr. T. Hiroi, who collected the RELAB spectra on our behalf, are greatly appreciated. This research is funded by the NASA Earth and Space Sciences Fellowship and by NASA Cosmochemistry grant NNX11AG27G and NASA Planetary Geology and Geophysics grant NNX10AJ57G.

46materials are preserved. Using rotationally resolved spectroscopy, especially in the mid-infrared, it may be possible to distinguish between these scenarios. JWST and the MIRI instrument will be able to help constrain the history of Minerva and similar asteroids related to the low metamorphic grade CO meteorites. A reliable detection of the 21-μm feature can be made with an SNR of ~100 if binning of the 5-28-μm region is employed.

The preservation of amorphous materials on the surface of Minerva and potentially other asteroids provide strong evidence that at least their exteriors have experienced minimal heating and aqueous alteration. Thus, asteroids rich in amorphous materials provide a unique window into the earliest epochs of the Solar System, prior to the onset of parent body processing, and offer an opportunity to explore the compositions and physical processes that occurred in the nebula and the early accretionary phases of asteroid formation.

**Acknowledgements:**

Spectra were acquired using the NASA Keck RELAB, a multiuser facility at Brown University. The efforts of Dr. T. Hiroi, who collected the RELAB spectra on our behalf, are greatly appreciated. This research is funded by the NASA Earth and Space Sciences Fellowship and by NASA Cosmochemistry grant NNX11AG27G and NASA Planetary Geology and Geophysics grant NNX10AJ57G.

**References**

Abreu, N. M., Brearley, A. J. (2010) Early solar system processes recorded in the matrices of two highly pristine CR3 carbonaceous chondrites, MET 00426 and QUE 99177. Geochimica et Cosmochimica Acta, Vol. 74, 1146-1171.

Abreu, N. M. (2016) Why is it so difficult to classify Renazzo-type (CR) Carbonaceous Chondrites? - Implications from TEM observations of matrices for the sequences of aqueous alteration. Geochimica et Cosmochimica Acta 194, p. 91-122.

Alexander, C. M. O'D., Bowden, R., Howard, K. T. (2014). A multi-technique search for the most primitive chondrites. Lunar and Planetary Science Conference, abstract 2667.

Alexander, C. M. O'D., Fogel, M., Yabuta, H., Cody, G. D. (2007) The origin and evolution of chondrites recorded in the elemental and isotopic compositions of their macromolecular organic matter. Geochimica et Cosmochimica Acta, 71, 4380-4403.